\documentclass{jnmp}

\usepackage{amsmath}

\JNMPnumberwithin{equation}{section}

\theoremstyle{definition}

\begin{document}

%
\renewcommand{\evenhead}{A Constantin and B Kolev}
\renewcommand{\oddhead}{Least Action Principle for an Integrable Shallow Water
Equation}

%
\thispagestyle{empty}

\FirstPageHead{8}{4}{2001}{\pageref{firstpage}--\pageref{lastpage}}{Letter}

\copyrightnote{2001}{A Constantin and B Kolev}

\Name{Least Action Principle for an Integrable Shallow Water
Equation}

\label{firstpage}

\Author{Adrian CONSTANTIN~$^\dag$ and Boris KOLEV~$^\ddag$}

\Address{$^\dag$ Department of Mathematics, Lund University, P.O. Box 118, SE-221 00 Lund, Sweden \\
~~E-mail: adrian.constantin@math.lu.se\\[10pt]
$^\ddag$ LATP, CMI Universit{\' e} de Provence,
    39 rue F. Joliot-Curie,
    13453 Marseille cedex 13,
    France \\
~~E-mail: kolev@cmi.univ-mrs.fr}

\Date{Received June 13, 2001;
Accepted July 26, 2001}

\begin{abstract}
\noindent For an integrable shallow water equation we describe a
geometrical approach showing that any two nearby fluid
configurations are successive states of a unique flow minimizing
the kinetic energy.
\end{abstract}

%
\section{Introduction}

The description of hydrodynamical flows by means of geodesics,
initiated by Arnold \cite{A}, allows one to apply the methods of
Riemannian geometry to the study of these flows. The geometric
approach consists mainly in formulating facts for the
infinite-dimensional case of fluid flows using results from
classical finite-dimensional Riemannian geometry cf. \cite{A-K}.
For the periodic Camassa-Holm \cite{C-H} equation, arising in the
study of shallow water waves, a rigorous geometric study can be
pursued \cite{C-K}. As a result, one obtains, for instance, a
proof of the fact that a state of the system can be transformed to
any nearby state by a unique flow. Of all possible paths joining
these two states, the system selects the one of minimal kinetic
energy - the Least Action Principle holds. In the next section we
highlight some aspects of the Camassa-Holm model and in Section 3
we present the geometric approach to the Least Action Principle
(the full details of which can be found in \cite{C-K}).

\section{The Camassa-Holm equation}

The spatially periodic equation
\begin{equation}\label{ch}
  u_t-u_{txx}+3uu_x=2u_xu_{xx}+uu_{xxx},
\end{equation}
in dimensionless space-time variables $(x,t)$ is a model for the
unidirectional propagation of two-dimensional shallow water waves
over a flat bottom \cite{C-H}, with $u(t,x)$ representing the
horizontal component of the velocity or, equivalently, the water's
free surface. The interest in periodic shallow water waves is
motivated by the observation that, as a matter of common
experience, the waves in a channel are approximately periodic. The
well-posedness issue (solutions exist, are unique and depend
continuously on the data), a perequisite to the usefulness of
(\ref{ch}) as a model equation for real wave phenomena, is settled
in \cite{H-M}. Equation (\ref{ch}) was already noticed in
\cite{F-F} as an abstract equation with infinitely many
conservation laws. This aspect is related to the fact that
(\ref{ch}) is an integrable infinite-dimensional Hamiltonian
system cf. \cite{C-M}: the equation can be converted into a
sequence of linear ordinary differential equations whose flows
have constant speed. Let us also note that equation (\ref{ch})
admits peaked travelling wave solutions that are solitons (two
travelling waves reconstitute their shape and size after
interacting with each other) cf. \cite{C-H}, and that the only way
a singularity can develop in a classical solution to (\ref{ch}) is
in the form of wave breaking \cite{C-E}: the solution remains
uniformly bounded while its slope becomes unbounded at a finite
time.

\section{The geometric approach}

From the Lagrangian viewpoint, each state of the system (\ref{ch})
is described by a diffeomorphism of the ambient space,
representing the rearrangement of the particles with respect to
their initial positions. The motion of the system is therefore a
path in the diffeomorphism group. Since a particle on the water's
free surface will always stay on the surface and (\ref{ch})
describes two-dimensional waves (no motion takes place in the
$y$-direction), we may regard the motion of (\ref{ch}) as that of
a one-dimensional membrane. In other words, the configuration
space of (\ref{ch}) can be reduced to the group $\mathcal{D}$ of
smooth orientation preserving diffeomorphisms of the circle. Let
$\varphi (t, \cdot )$ be a path in the diffeomorphism group,
starting at the identity $\varphi (0, x) = x$,  $x \in \mathbb{S}$
$(\mathbb{S}$ being the unit circle), representing the evolution
of an initial state for (\ref{ch}). The material velocity field is
$(t,x)\mapsto \varphi_t (t,x)$ while the spatial velocity field is
$u(t,x) = \varphi_t (t,\varphi^{-1}(t,x))$. In terms of $u$, we
have the Eulerian description (from the viewpoint of a fixed
observer) while $\varphi_t$ represents the Lagrangian viewpoint
(the motion as seen by following each particle): for a fluid
particle initially located at $x$, $\varphi_t (t,x)$ is its
velocity at time $t$, while $u(t,\varphi (t,x))$ is the velocity
at the location $\varphi (t,x)$.

In Lagrangian mechanics, the motion of a mechanical system is a
critical point of a certain functional (the Action), defined on
all the paths in configuration space having fixed endpoints. For a
system with no external forces (inertial system), this Action is
precisely the kinetic energy. To the order of approximation to
which (\ref{ch}) was obtained from the governing equations for
water waves, $u(t,x)$ represents the horizontal velocity component
and $u_x(t,x)$ is the vertical velocity component at the free
surface cf. \cite{H}. Hence, the kinetic energy on the free
surface at instant $t$ is
\begin{equation}\label{k0}
\frac{1}{2}\,\int_{\mathbb{S}} \Bigl(u^2(t,x) + u^2_x(t,x)\Bigr)\,
dx.
\end{equation}
The fact that there is no preferred initial state in the
configuration space of (\ref{ch}) suggests a right invariance
property: observe that if we replace the path $t\mapsto \varphi
(t, \cdot )$ by $t\mapsto \varphi (t,\varphi_0 ( \cdot ) )$ for a
fixed time-independent $\varphi_0 \in \mathcal{D}$, then the
spatial velocity $u$ is unchanged. Hence, the kinetic energy of a
state $\varphi \in \mathcal{D}$ of the system (\ref{ch}) is given
by
\begin{equation}\label{k}
K(\varphi, \varphi_t) = \frac{1}{2}\,\int_{\mathbb{S}}
\Bigl\{(\varphi_t\circ \varphi^{-1})^2 + [\partial_x\,(\varphi_t
\circ\varphi^{-1})]^2\Bigr\}\, dx.
\end{equation}
In conclusion, the motion is described by the critical points of
the Action
\begin{equation}\label{a}
\int_0^T K(\varphi, \varphi_t) \, dt = \frac{1}{2}\,\int_0^T
\int_{\mathbb{S}} \Bigl\{(\varphi_t\circ \varphi^{-1})^2 +
[\partial_x\,(\varphi_t \circ\varphi^{-1})]^2\Bigr\}\, dx\, dt,
\end{equation}
over the path $\{\varphi (t, \cdot );\, t\in [0,T]\}$ in
$\mathcal{D}$. Note that the quadratic functional (\ref{k})
defines a right-invariant Riemannian metric on the Lie group
$\mathcal{D}$ and the geodesic flow of this metric corresponds to
the critical points of the Action. The geodesic curve $t\mapsto
\varphi (t, \cdot )$ in $\mathcal{D}$ starting at the identity
$\varphi (0,\cdot ) = Id$ in the direction $u_0 \in  C^{\infty}
(\mathbb{S})$ satisfies the equation
\begin{equation}\label{g}
\begin{cases}
    \varphi_t = v\\
    v_t = P (\varphi, v),
  \end{cases}
\end{equation}
where the operator $P$ is given by
\begin{equation}\label{P}
  P(\varphi, v ) = - \Bigl\{\partial_x \, (1 - \partial_x^2 )^{-1}
  \Bigl( (v\circ \varphi^{-1})^2 + \frac{1}{2}\  [(v\circ
  \varphi^{-1})_x]^2\Bigr)\Bigr\}\circ \varphi.
\end{equation}
In terms of the Eulerian velocity $u = \varphi_t \circ
\varphi^{-1}$, the geodesic equation (\ref{g}) is precisely the
Camassa-Holm equation (\ref{ch}) as pointed out in \cite{M}. This
resembles the fact that the Euler equation in fluid mechanics is
an expression of the geodesic flow in the group of incompressible
diffeomorphisms cf. \cite{A}, \cite{E-M}.

It is possible to re-express certain ideal fluid flows, as well as
some geophysical fluid flows, as geodesic flows on their
configuration spaces cf. \cite{A-K} but the results obtained are
formal in character. It turns out that for the Camassa-Holm
equation a rigorous geometric study can be pursued. Without going
into the details of the intricate analysis made in \cite{C-K}, let
us present the results obtained. A study of the geodesic equation
(\ref{g}) yields the local existence of the geodesic flow on
$\mathcal{D}$. Interestingly, some geodesics are defined globally
in time, while on certain geodesic paths the diffeomorphisms
flatten out in finite time. Pursuing the analysis, one can prove
that a geodesic on $\mathcal{D}$ is locally the shortest path
between two nearby elements of $\mathcal{D}$ - the Least Action
Principle holds for the Camassa-Holm equation (\ref{ch}). This
result has a direct physical interpretation and illustrates the
power of the geometric approach. It simply means that a state of
the system (\ref{ch}) is transformed to another nearby state by
going through a uniquely determined flow that minimizes the
kinetic energy. To the best of our knowledge, this is the first
example where the Riemannian geometric approach can be pursued
rigorously to obtain physically relevant information. Previous
studies introduce enlarged configuration spaces \cite{E-M} where
the geodesics are only formal objects aimed to provide insight
into the actual configuration space; however, the passage from
these intermediate spaces to the configuration space remains an
open question. Other approaches circumvent the analytical
difficulties encountered in working with the configuration space
by defining generalized flows (where particles may split and
collide) but the physical meaning of these generalized flows is
not understood cf. \cite{A-K}.

\label{lastpage}

\end{document}